# MULTIPLE LINEAR REGRESSION AND CORRELATION:
# A GEOMETRIC ANALYSIS

**Abstract**

In this review article we consider linear regression analysis from a geometric perspective, looking at standard methods and outputs in terms of the lengths of the relevant vectors and the angles between these vectors. We show that standard regression output can be written in terms of the lengths and angles between the various input vectors, such that this geometric information is sufficient in linear regression problems. This allows us to obtain a standard formula for multiple correlation and give a geometric interpretation to this. We examine how multicollinearity affects the total explanatory power of the data, and we examine a counter-intuitive phenomena called "enhancement" where the total information from the explanatory vectors is greater than the sum of the marginal parts.

MULTIPLE LINEAR REGRESSION; CORRELATION; GEOMETRIC ANALYSIS.

# MULTIPLE LINEAR REGRESSION AND CORRELATION: A GEOMETRIC ANALYSIS

BEN O'NEILL[*], *Australian National University*[**]

29 MARCH 2019

Geometric methods using vectors have been pressed into service in the exposition of statistical methods in a wide range of applications. An exposition of early work can be found in Herr (1980) showing that geometric methods were understood by eminent statisticians such as Fisher, Bartlett, Durbin and Kendall. Later work in Bryant (1984) and Saville and Wood (1986) gave a more general exposition of statistical procedures from a geometric perspective. Nowadays there are entire texts on the subject, such as Saville and Wood (1991) and Wickens (1995), giving practitioners a good exposition of results in this field. In this paper we continue this tradition by looking at multiple linear regression from a geometric perspective and showing how educators in this field can present multiple regression as an extension of basic correlation analysis. This is designed to assist teachers of regression analysis to impart the core results of the subject using a small number of simple vector-based geometric tools.

Linear regression models form a baseline for the analysis of multivariate data, and are applicable in a wide range of statistical problems (for discussion see Cohen 1968). The subject is a core component of statistical education and most universities have one or more courses devoted to the subject. The vector-based geometric approach is a useful method of analysis for linear regression, requiring an understanding of matrix algebra, up to the level of eigenvectors, eigenvalues and the spectral decomposition.

In this review article we develop a geometric representation of linear regression that has various advantages in interpretation and implementation. Geometric framing of the regression model explains the relationship between the coefficient of determination and the pairwise correlations between vectors, elucidating the problem of multicollinearity. It also allows us to see the minimal information that impacts on the "goodness of fit" of the model, through the coefficient of determination. The geometric framing of the model also assists with model selection, since it allows practitioners to rapidly calculate the coefficient of determination for subsets of explanatory variables without calculating intermediary regression quantities. Analysis of the coefficient of determination along geometric lines also alerts practitioners to some counter-intuitive properties of linear regression involving "enhancement" effects, and provides a natural introduction into principal component analysis.

---

[*] E-mail addresses: ben.oneill@hotmail.com; ben.oneill@anu.edu.au.
[**] Research School of Population Health, Australian National University, Canberra ACT 0200.

**1. Estimated correlation between regression variables**

It is standard in linear regression analysis to look at the correlation between variables as an exploratory step prior to implementation of the model (Hocking 1996, pp. 208-212). This exercise also makes a good instructional precursor for regression analysis in introductory courses. Introductory courses on regression analysis usually follow courses in which students have been exposed to pairwise correlation and scatterplots of variables. So long as the number of variables is not too large, this usually involves the creation of a scatterplot matrix showing all pairs of variables with corresponding correlation estimates for each pair. Sometimes this is supplemented with analysis of three-dimensional scatterplots for triplets of variables. In cases where the number of variables is large there are other graphical techniques which can be used (see e.g., Hills 1969, Corsten and Gabriel 1976, Murdoch and Chow 1996, and Trosset 2005).

In any case, any form of this kind of preliminary analysis requires correlation estimates for all the pairs of variables. The standard multiple linear regression model involves a response vector and a set of corresponding explanatory vectors:

$$\boldsymbol{y} = \begin{bmatrix} y_1 \\ y_2 \\ \vdots \\ y_n \end{bmatrix} \quad \boldsymbol{x}_1 = \begin{bmatrix} x_{1,1} \\ x_{2,1} \\ \vdots \\ x_{n,1} \end{bmatrix} \quad \cdots \quad \boldsymbol{x}_m = \begin{bmatrix} x_{1,m} \\ x_{2,m} \\ \vdots \\ x_{n,m} \end{bmatrix}.$$

This designation corresponds to observation of $n$ data points, consisting of a response variable and $m$ explanatory variables. To incorporate an intercept term into the model we will assume that the response vector and explanatory vectors are mean-adjusted ("centred") to zero (i.e., $\sum y_i = \sum x_{i,1} = \cdots = \sum x_{i,m} = 0$).[1]

---

[1] This is a harmless simplification that we will discuss further when we introduce linear regression (see e.g., Davidson and MacKinnon 1993, pp. 19-24).

The product-moment correlation between the various pairs of regression variables can be determined by simple vector calculations. Since the vectors are already assumed to have already been mean-adjusted to zero we have:

$$R_i = \text{corr}(\boldsymbol{y}, \boldsymbol{x}_i) = \frac{\boldsymbol{x}_i \cdot \boldsymbol{y}}{\|\boldsymbol{x}_i\|\|\boldsymbol{y}\|} \qquad R_{i,k} = \text{corr}(\boldsymbol{x}_i, \boldsymbol{x}_k) = \frac{\boldsymbol{x}_i \cdot \boldsymbol{x}_k}{\|\boldsymbol{x}_i\|\|\boldsymbol{x}_k\|}.$$

The estimated correlation values can be summarized as a correlation matrix:

$$\boldsymbol{\Phi} = \textbf{corr}(\boldsymbol{y}, \boldsymbol{x}_1, \ldots, \boldsymbol{x}_m) = \begin{bmatrix} 1 & R_1 & R_2 & \cdots & R_m \\ R_1 & 1 & R_{2,1} & \cdots & R_{m,1} \\ R_2 & R_{1,2} & 1 & \cdots & R_{m,2} \\ \vdots & \vdots & \vdots & \ddots & \vdots \\ R_m & R_{1,m} & R_{2,m} & \cdots & 1 \end{bmatrix}.$$

There are substantive restrictions on the elements of a correlation matrix. The matrix is symmetric non-negative definite with unit diagonals. This means that the matrix is diagonalizable and has non-negative eigenvalues with orthogonal eigenvectors. (If the input vectors are linearly independent then the matrix is positive definite, so that the eigenvalues are all strictly positive.) Moreover, since the sum of the eigenvalues is the trace of the matrix, the eigenvalues of a correlation matrix have unit average. Any principal submatrix of a correlation matrix is also a correlation matrix, so the required properties must hold over all principal submatrices. Any matrix with these properties is a valid correlation matrix.

Since our analysis will take a geometric focus, it is useful for us to write the correlation matrix in geometric terms. To do this, we note that the product-moment correlations can also be represented geometrically in terms of the angles between the pairs of vectors using standard results for vector angles:

$$R_i = \cos \omega_i \qquad \omega_i = \text{Smallest angle between } \boldsymbol{y} \text{ and } \boldsymbol{x}_i,$$
$$R_{i,k} = \cos \theta_{i,k} \qquad \theta_{i,k} = \text{Smallest angle between } \boldsymbol{x}_i \text{ and } \boldsymbol{x}_k.$$

(For discussion of this relationship see e.g., Rodgers and Nicewander 1988, pp. 61, 63.) Hence, the correlation matrix can be rewritten as a matrix of cosines of angles between the vectors in the regression analysis (the unit values on the diagonal can be recognised as the cosine of the angle of each vector with itself):

$$\boldsymbol{\Phi} = \textbf{corr}(\boldsymbol{y}, \boldsymbol{x}_1, \ldots, \boldsymbol{x}_m) = \begin{bmatrix} 1 & \cos \omega_1 & \cos \omega_2 & \cdots & \cos \omega_m \\ \cos \omega_1 & 1 & \cos \theta_{1,2} & \cdots & \cos \theta_{1,m} \\ \cos \omega_2 & \cos \theta_{1,2} & 1 & \cdots & \cos \theta_{2,m} \\ \vdots & \vdots & \vdots & \ddots & \vdots \\ \cos \omega_m & \cos \theta_{1,m} & \cos \theta_{2,m} & \cdots & 1 \end{bmatrix}.$$

For pedagogical purposes, we note that geometric analysis of correlation can be shown using a combination of algebraic calculations and pictures of vectors in a vector space. It is useful to visualise correlation in terms of angles between vectors of data values. Rodgers and Nicewater (1988) and Rovine and von Eye (1997) give some interesting interpretations for the estimated correlation coefficient, including some interpretations making reference to univariate regression analysis. Taking raw data and using this to construct a scatterplot matrix with correlation values is a good way to show correlation results between pairs of variables. Students of regression can be invited to speculate on the multivariate interactions in the data and later compare their speculations with the results from a proper multiple regression analysis.

**EXAMPLE 1:** We consider data taken from Thomas (1990) which shows mortality data in small American cities. The response variable of interest is the death rate in each city and there are four prospective explanatory variables. The scatterplot matrix for this data with accompanying correlation estimates is shown in Figure 1. Students could be asked to construct this kind of plot from raw data and invited to speculate on the relationships in the data based on the scatterplot and correlation results, if only to demonstrate the difficulty in seeing multivariate interactions in this kind of plot. ∎

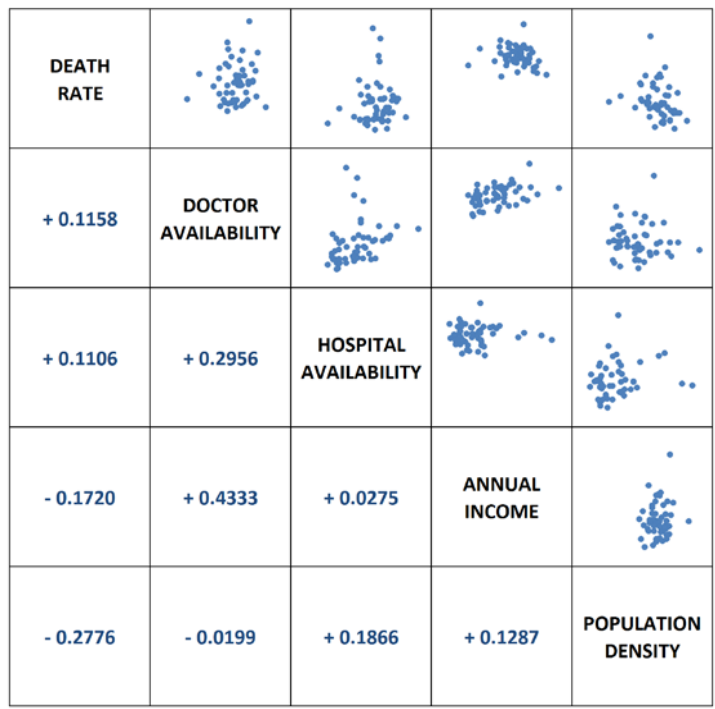

**FIGURE 1:** Scatterplot matrix for mortality data in Thomas (1990)

## 2. The multiple linear regression model

In multiple linear regression models the conditional expectation of the response vector is considered to be a linear function of the explanatory vectors, with deviation from this conditional expectation specified by an error vector. This is based on a design matrix, and slope coefficient and error vectors given respectively by:

$$x = [x_1 \cdots x_m] = \begin{bmatrix} x_{1,1} & x_{1,2} & \cdots & x_{1,m} \\ x_{2,1} & x_{2,2} & \cdots & x_{2,m} \\ \vdots & \vdots & \ddots & \vdots \\ x_{n,1} & x_{n,2} & \cdots & x_{n,m} \end{bmatrix} \quad \boldsymbol{\beta} = \begin{bmatrix} \beta_1 \\ \beta_2 \\ \vdots \\ \beta_m \end{bmatrix} \quad \boldsymbol{\varepsilon} = \begin{bmatrix} \varepsilon_1 \\ \varepsilon_2 \\ \vdots \\ \varepsilon_n \end{bmatrix}.$$

(Note that the intercept parameter is omitted here since the vectors have already been mean-adjusted to zero.) The linear regression model is defined by the equations:

$$Y = x\boldsymbol{\beta} + \boldsymbol{\varepsilon} \qquad \mathbb{E}(\boldsymbol{\varepsilon}|x) = \mathbf{0} \qquad \mathbb{V}(\boldsymbol{\varepsilon}|x) = \sigma^2 I.$$

We take the errors to be homoscedastic and uncorrelated. (In the Gaussian regression model the error vector is also assumed to be normally distributed.) Instructors may wish to show the corresponding scalar equations for each data point and accompany this by standard graphs of data points scattered around an unseen regression line.

We fit the model by estimating the unknown parameter vector $\boldsymbol{\beta}$ using ordinary least-squares (OLS) estimation. This is known to give the best-linear-unbiased-estimator for the coefficient vector and is also the MLE in the case of the Gaussian linear regression model (Hocking 1996, pp. 77-78). We assume that the vectors $x_1, \ldots, x_m$ are linearly independent, which ensures that the correlation matrix of the explanatory vectors is non-singular. OLS estimation gives us well-known forms for the estimated coefficient vector, hat matrix and annihilator matrix:[2]

$$\hat{\boldsymbol{\beta}} = (x^T x)^{-1}(x^T y) \qquad h = x(x^T x)^{-1} x^T \qquad a = I - h.$$

Although it is possible to proceed with explicit inclusion of an intercept term (without mean-adjusting the other input vectors) we simplify this by incorporating the intercept implicitly into the specification of the input vectors by taking these to be mean-adjusted to zero (see e.g., Davidson and MacKinnon 1993, pp. 19-24). The intercept parameter can easily be added into the model *post hoc* if needed.[3]

---

[2] For completeness, it is also worth noting that the matrix $x^T x$ is the Gramian matrix of the design matrix and the matrix $(x^T x)^{-1} x^T$ is the Moore-Penrose pseudo-inverse of the design matrix.

[3] If $\bar{y}$ and $\bar{x}_1, \ldots, \bar{x}_m$ are the means of the unadjusted vectors in the regression then the estimated intercept parameter is $\hat{\beta}_0 = \bar{y} - \sum_{k=1}^{m} \hat{\beta}_k \bar{x}_k$.

The OLS estimator minimises the norm of the residual vector in the regression. It can be derived using vector calculus as the solution of the optimisation corresponding to the MLE, or it can be derived using linear algebra as the projection of the response vector onto the column space of the design matrix. (The hat matrix is the projection matrix which maps the response onto to the predicted response.) The response vector is then decomposed as $\boldsymbol{y} = \hat{\boldsymbol{y}} + \boldsymbol{r}$ with predicted response vector $\hat{\boldsymbol{y}} = \boldsymbol{x}\hat{\boldsymbol{\beta}} = \boldsymbol{h}\boldsymbol{y}$ and residual vector $\boldsymbol{r} = \boldsymbol{y} - \hat{\boldsymbol{y}} = \boldsymbol{a}\boldsymbol{y}$. Students should be encouraged to visualise the vector projection geometrically using a three-dimensional picture of a larger vector space.

Since the effect of the intercept term has already been filtered out of the model,[4] the sums-of-squares, degrees-of-freedom, and mean-squares in the Analysis of Variance (ANOVA) are given by:

$$SS_{Tot} = \|\boldsymbol{y}\|^2 \qquad df_{Tot} = n - 1 \qquad MS_{Tot} = \frac{\|\boldsymbol{y}\|^2}{n-1},$$

$$SS_{Reg} = \|\hat{\boldsymbol{y}}\|^2 \qquad df_{Reg} = m \qquad MS_{Reg} = \frac{\|\hat{\boldsymbol{y}}\|^2}{m},$$

$$SS_{Res} = \|\boldsymbol{r}\|^2 \qquad df_{Res} = n - m - 1 \qquad MS_{Res} = \frac{\|\boldsymbol{r}\|^2}{n-m-1}.$$

The other ANOVA statistics are:

$$\hat{\sigma}_Y^2 = MS_{Tot} = \frac{\|\boldsymbol{y}\|^2}{n-1} \qquad \hat{\sigma}^2 = MS_{Res} = \frac{\|\boldsymbol{r}\|^2}{n-m-1},$$

$$R^2 = \frac{SS_{Reg}}{SS_{Tot}} = \left(\frac{\|\hat{\boldsymbol{y}}\|}{\|\boldsymbol{y}\|}\right)^2 \qquad F = \frac{MS_{Reg}}{MS_{Res}} = \frac{n-m-1}{m}\left(\frac{\|\hat{\boldsymbol{y}}\|}{\|\boldsymbol{r}\|}\right)^2.$$

All of these quantities can be explained in terms of the decomposition of the response vector into the predicted and residual parts. Visualisation of the vector projection should make it easy to see the Pythagorean result that $\|\boldsymbol{y}\|^2 = \|\hat{\boldsymbol{y}}\|^2 + \|\boldsymbol{r}\|^2$ as well as the effects of applying the hat matrix or annihilator matrix to regression vectors. Degrees-of-freedom can be given their proper technical meaning as the dimensions of the vector subspaces involved in the decomposition, but some instructors may prefer to give a more heuristic explanation in terms of effective sample-size.

---

[4] It is important to note that the formulae here are based on a regression model with an intercept term that has been incorporated implicitly by mean-adjustment of the input vectors. To obtain the formulae for a model without an intercept term all that needs to occur is to refrain from mean-adjustment of the input vectors and adjust the total and residual degrees-of-freedom accordingly.

## 3. Geometric analysis of the multiple linear regression model

The above exposition of correlation and linear regression is based on direct use of the input vectors. The pure geometric approach to regression analysis instead looks at the input vectors based on their lengths and their angles with respect to each other. From our response vector and explanatory vectors we define the normed vectors:

$$\acute{y} \equiv \frac{y}{\|y\|} \qquad \acute{x}_1 \equiv \frac{x_1}{\|x_1\|} \qquad \cdots \qquad \acute{x}_m \equiv \frac{x_m}{\|x_m\|}.$$

So as not to lose information, we arrange the lengths of the explanatory vectors into a norming matrix, which we denote using a slight abuse of notation[5] as:

$$\|x\| \equiv \begin{bmatrix} \|x_1\| & 0 & \cdots & 0 \\ 0 & \|x_2\| & \cdots & 0 \\ \vdots & \vdots & \ddots & \vdots \\ 0 & 0 & \cdots & \|x_m\| \end{bmatrix}.$$

We define the normed design matrix by $\acute{x} \equiv x\|x\|^{-1} = [\acute{x}_1 \ \cdots \ \acute{x}_m]$. We define the **goodness of fit vector** and **design correlation matrix** respectively by:

$$\Omega \equiv \begin{bmatrix} R_1 \\ R_2 \\ \vdots \\ R_m \end{bmatrix} \qquad \Theta \equiv \begin{bmatrix} 1 & R_{2,1} & \cdots & R_{m,1} \\ R_{1,2} & 1 & \cdots & R_{m,2} \\ \vdots & \vdots & \ddots & \vdots \\ R_{1,m} & R_{2,m} & \cdots & 1 \end{bmatrix}.$$

(The reason for the latter name should be obvious. The reason for the former will become evident later.) We can then decompose the previous correlation matrix as:

$$\Phi = \begin{bmatrix} 1 & \Omega^{\mathrm{T}} \\ \Omega & \Theta \end{bmatrix}.$$

(The matrix $\Theta$ is the principle submatrix of $\Phi$ obtained by removing the response.) These norms and matrices allow us to encapsulate all the geometric information in our input vectors and separate that information into two distinct parts — the objects $\|y\|$ and $\|x\|$ which refer only to the lengths of the input vectors, and the objects $\Omega$ and $\Theta$ which refer to the angles between the input vectors, or equivalently, to their product-moment correlations.

---

[5] Here we refer to the norming matrix using notation for a norm of the design matrix. Strictly speaking, this is not the norm of a matrix — for one thing it is a matrix itself and not a scalar. Nevertheless, we think this notation is useful in the present context, to make clear the demarcation between the objects referring to the length of the vectors, and objects referring to their angle. The reader should be careful to remember that the norming matrix is a matrix, not a scalar, and so it is not commutative with other objects in the formulae we put forward here. Hopefully more is gained by our abuse of notation than is lost with this potential confusion.

It turns out that it is possible to express all the quantities in standard linear regression output in terms of these geometric objects. In particular, it is easy to show that:

$$\acute{x}^T \acute{x} = \Theta \qquad \acute{x}^T \acute{y} = \Omega,$$
$$x^T x = \|x\| \Theta \|x\| \qquad x^T y = \|y\| \|x\| \Omega.$$

With a little algebra this gives us:

$$\widehat{\boldsymbol{\beta}} = (\|y\| \|x\|^{-1})(\Theta^{-1} \Omega) \qquad h = \acute{x} \Theta^{-1} \acute{x}^T \qquad a = I - \acute{x} \Theta^{-1} \acute{x}^T.$$

We also have the decomposition $\acute{y} = \widehat{\acute{y}} + \acute{r}$ where predicted normed response vector is $\widehat{\acute{y}} = h\acute{y} = \acute{x}(\Theta^{-1} \Omega)$ and the normed residual vector is $\acute{r} = \acute{y} - \widehat{\acute{y}} = a\acute{y}$. The sum-of-squares and mean-square statistics are:

$$SS_{Tot} = \|y\|^2 \qquad\qquad MS_{Tot} = \frac{\|y\|^2}{n-1},$$

$$SS_{Reg} = \|y\|^2 \Omega^T \Theta^{-1} \Omega \qquad\qquad MS_{Reg} = \frac{\|y\|^2}{m} \Omega^T \Theta^{-1} \Omega,$$

$$SS_{Res} = \|y\|^2 (1 - \Omega^T \Theta^{-1} \Omega) \qquad\qquad MS_{Res} = \frac{\|y\|^2}{n-m-1}(1 - \Omega^T \Theta^{-1} \Omega).$$

The other ANOVA statistics are:

$$\hat{\sigma}_Y^2 = \frac{\|y\|^2}{n-1} \qquad\qquad \hat{\sigma}^2 = \frac{\|y\|^2}{n-1}(1 - \Omega^T \Theta^{-1} \Omega),$$

$$R^2 = \Omega^T \Theta^{-1} \Omega \qquad\qquad F = \frac{n-m-1}{m} \frac{\Omega^T \Theta^{-1} \Omega}{1 - \Omega^T \Theta^{-1} \Omega}.$$

This shows us that the ANOVA quantities are determined entirely by the lengths of the input vectors and the angles between these vectors. From these formulae we can see that the coefficient of determination and F-statistic are scale-free outputs.

All of this is intuitively reasonable and corresponds with what we would expect when we visualise the effect of uniform scaling and rotation on input vectors in a vector space. If all the input vectors are scaled by some constant multiple then they retain their angles and relationships and the only change we would expect is a corresponding effect on scale-dependent outputs like the coefficient estimates, sums-of-squares and mean-squares. Similarly, if all of the input vectors are rotated in some way inside the vector space we expect that this will have no effect on any of the relations between them, and so it should not affect any of the regression outputs. Students should be encouraged to visualise these intuitive results in geometric terms. Pictures of vectors in a vector space can help to assist instruction of students.

The above results allow us to implement goodness-of-fit tests for linear regression based solely on the correlations between the pairs of input vectors. Hence, we see that the correlation values shown in the scatterplot matrix in Figure 1 are actually sufficient to test whether or not there is evidence of a relationship between the response and the set of explanatory variables. This may be contrary to intuition since some practitioners assume that the pairwise correlation values would not be sufficient for a holistic analysis of multiple linear regression.

It is a useful exercise for students to implement regression analysis on the original set of regression vectors $y, x_1, ..., x_m$ and then repeat the same analysis using the pure geometric analysis which uses the normed vectors $\acute{y}, \acute{x}_1, ..., \acute{x}_m$. This can be done to show in practical terms that the results of the geometric analysis are valid.

**EXAMPLE 2:** Continuing Example 1 we are now able to use the correlation matrix and the lengths of the input vectors to obtain the full regression output. It is easily shown that direct implementation of the regression model on the input vectors yields the same output as the above formulae based on their geometric properties. We have:

$$\Theta = \begin{bmatrix} 1.0000 & 0.2956 & 0.4333 & -0.0199 \\ 0.2956 & 1.0000 & 0.0275 & 0.1866 \\ 0.4333 & 0.0275 & 1.0000 & 0.1287 \\ -0.0199 & 0.1866 & 0.1287 & 1.0000 \end{bmatrix} \quad \Omega = \begin{bmatrix} 0.1158 \\ 0.1106 \\ -0.1720 \\ -0.2776 \end{bmatrix}.$$

The coefficient of determination is:

$$R^2 = \Omega^T \Theta^{-1} \Omega = 0.1437.$$

There are $n = 53$ data points and $m = 4$ explanatory variables giving F-statistic:

$$F = \frac{53 - 4 - 1}{4} \cdot \frac{0.1437}{1 - 0.1437} = 12 \cdot \frac{0.1437}{0.8563} = 2.0138 \qquad p = 0.1075.$$

Hence, in this case there is no evidence of a linear relationship between the response variable and the set of explanatory variables. ∎

**4. The coefficient of determination and multiple correlation**

If we would like to relate the situation of two random variables with the situation of multiple explanatory variables then the statistic of most interest to us is the coefficient of determination. In the two-variable case the strength of the linear relationship between the variables can be measured by the square of the correlation coefficient.[6] The coefficient of determination is a multivariate extension of this measure — it measures the strength of the linear relationship between the response variable and the set of explanatory variables.

The square root of the coefficient of determination gives us the multiple correlation coefficient, which is a multivariate extension of the absolute correlation (Abdi 2007). This statistic measures the total absolute correlation between a set of explanatory variables and the response variable in a regression. It can be used to look at the interaction between the various explanatory variables and the effects of their inclusion or exclusion from the model. In fact, there are several works on regression that focus on its connection with pairwise correlation measures and the resulting framework for analysing interactions between explanatory variables (see e.g., Aiken and West 1991, Belsley 1991, Cohen et al 2003).

From the above geometric analysis we can see that the coefficient of determination in our analysis is given by $R^2 = \boldsymbol{\Omega}^\mathrm{T} \boldsymbol{\Theta}^{-1} \boldsymbol{\Omega}$ (see e.g., Mardia, Kent and Bibby 1979, p. 168). This expression is a quadratic form with coefficient matrix given by the inverse of the design correlation matrix and argument vector given by the goodness-of-fit vector. It is a basic expression used in texts which analyse the relationship between correlation and multiple regression. It is trivial to show that in the case of a single explanatory variable the coefficient of determination is the square of the correlation between the response and explanatory vectors. (This explains why we refer to the vectors of these correlations as the goodness-of-fit vector.) If the explanatory vectors are orthogonal then the correlation between them is zero and we have $\boldsymbol{\Theta} = \boldsymbol{I}$ so that $R^2 = \|\boldsymbol{\Omega}\|^2 = \sum R_i^2$. Intuitively, this is the case where the explanatory variables are giving distinct information, so that the total information is the sum of the parts.

---

[6] The raw correlation coefficient captures the strength and direction of the linear relationship. If we take the absolute value we remove the directionality and we obtain a statistic measuring only the strength of the relationship. Any monotonic transformation of this statistic preserves this measure, and hence, the squared correlation coefficient does equally well as a measure of strength.

**Two explanatory variables:** The case where $m = 2$ is interesting, and has been used in examples in much of the literature looking at the coefficient of determination. It is a simple case for students to consider when learning the subject. In this case we have:

$$\mathbf{\Omega} = \begin{bmatrix} R_1 \\ R_2 \end{bmatrix} \qquad \mathbf{\Theta} = \begin{bmatrix} 1 & R_{1,2} \\ R_{1,2} & 1 \end{bmatrix} \qquad \mathbf{\Theta}^{-1} = \frac{1}{1 - R_{1,2}^2} \begin{bmatrix} 1 & -R_{1,2} \\ -R_{1,2} & 1 \end{bmatrix}.$$

This means that:

$$R^2 = \mathbf{\Omega}^\mathrm{T} \mathbf{\Theta}^{-1} \mathbf{\Omega} = \frac{1}{1 - R_{1,2}^2} [R_1 \quad R_2] \begin{bmatrix} 1 & -R_{1,2} \\ -R_{1,2} & 1 \end{bmatrix} \begin{bmatrix} R_1 \\ R_2 \end{bmatrix}$$

$$= \frac{1}{1 - R_{1,2}^2} \left( R_1^2 + R_2^2 - 2 R_{1,2} R_1 R_2 \right).$$

This is a special case of our more general formula. The present formula is given in Hamilton (1987) (p. 130, Eq 8) and has been used by later authors to examine the relationship between the coefficient of determination and correlation. ∎

We can obtain a geometric characterisation of the coefficient of determination by using the spectral decomposition of the design correlation matrix. The explanatory vectors are assumed to be linearly independent, so this is a positive definite matrix. This means that it has positive eigenvalues ordered by $\lambda_1 \geq \cdots \geq \lambda_m$ with $\sum \lambda_i = m$ and it has corresponding orthogonal unit eigenvectors $\boldsymbol{v}_1, \dots, \boldsymbol{v}_m$ (normed to unit length so that the eigenvector matrix is orthonormal). These eigenvectors represent the vector axes for the principal components of the standardised explanatory variables.

The above geometric framework for regression analysis allows us to decompose the coefficient of determination into a sum of parts representing the contribution from each of the principal components of the design matrix. To see this, we note that the matrix $\mathbf{\Theta}$ can be expressed by its spectral decomposition $\mathbf{\Theta} = \boldsymbol{v} \mathbf{\Lambda} \boldsymbol{v}^\mathrm{T}$ where:

$$\boldsymbol{v} \equiv [\boldsymbol{v}_1 \quad \boldsymbol{v}_2 \quad \cdots \quad \boldsymbol{v}_m] \qquad \mathbf{\Lambda} \equiv \begin{bmatrix} \lambda_1 & 0 & \cdots & 0 \\ 0 & \lambda_2 & \cdots & 0 \\ \vdots & \vdots & \ddots & \vdots \\ 0 & 0 & \cdots & \lambda_m \end{bmatrix}.$$

We let $\boldsymbol{z} = \acute{\boldsymbol{x}} \boldsymbol{v}$ be the matrix of principal components of the standardised data. Each vector $\boldsymbol{z}_k = \acute{\boldsymbol{x}} \boldsymbol{v}_k$ is the principal component corresponding to the eigenvector $\boldsymbol{v}_k$. (These principal components are orthogonal with $\|\boldsymbol{z}_k\|^2 = \lambda_k$.) We then have:

$$S_k \equiv \mathrm{corr}(\boldsymbol{y}, \boldsymbol{z}_k) = \frac{\boldsymbol{y} \cdot \boldsymbol{z}_k}{\|\boldsymbol{y}\| \|\boldsymbol{z}_k\|} = \frac{\boldsymbol{z}_k^\mathrm{T} \acute{\boldsymbol{y}}}{\sqrt{\lambda_k}} = \frac{\boldsymbol{v}_k^\mathrm{T} \acute{\boldsymbol{x}}^\mathrm{T} \acute{\boldsymbol{y}}}{\sqrt{\lambda_k}} = \frac{\boldsymbol{v}_k^\mathrm{T} \mathbf{\Omega}}{\sqrt{\lambda_k}} = \frac{\mathbf{\Omega} \cdot \boldsymbol{v}_k}{\sqrt{\lambda_k}}.$$

We therefore have:
$$R^2 = \boldsymbol{\Omega}^\mathrm{T}\boldsymbol{\Theta}^{-1}\boldsymbol{\Omega} = \boldsymbol{\Omega}^\mathrm{T}(\boldsymbol{\upsilon}\boldsymbol{\Lambda}\boldsymbol{\upsilon}^\mathrm{T})^{-1}\boldsymbol{\Omega} = (\boldsymbol{\upsilon}^\mathrm{T}\boldsymbol{\Omega})^\mathrm{T}\boldsymbol{\Lambda}^{-1}(\boldsymbol{\upsilon}^\mathrm{T}\boldsymbol{\Omega})$$
$$= \sum_{k=1}^{m} \frac{1}{\lambda_k}(\boldsymbol{\Omega}\cdot\boldsymbol{\upsilon}_k)^2 = \sum_{k=1}^{m} S_k^2.$$

We can see from this characterisation of the coefficient of determination that it can be considered as the sum of $m$ parts, where each part represents the contribution from the principal component for $\boldsymbol{\upsilon}_k$. The coefficient is the sum-of-squares of the pairwise correlations between the vectors $\boldsymbol{z}_k$ and the response variable $\boldsymbol{y}$.

**Two explanatory variables (alternative):** In the case where $m = 2$ the eigenvalue and eigenvector matrices for the design correlation matrix are given respectively by:

$$\boldsymbol{\Lambda} = \begin{bmatrix} 1-R_{12} & 0 \\ 0 & 1+R_{12} \end{bmatrix} \qquad \boldsymbol{\upsilon} = \begin{bmatrix} 1/\sqrt{2} & 1/\sqrt{2} \\ -1/\sqrt{2} & 1/\sqrt{2} \end{bmatrix}.$$

We therefore have:
$$S_1 = \frac{\boldsymbol{\Omega}\cdot\boldsymbol{\upsilon}_1}{\sqrt{\lambda_1}} = \frac{1}{\sqrt{2}}\cdot\frac{R_1-R_2}{\sqrt{1-R_{12}}} \qquad S_2 = \frac{\boldsymbol{\Omega}\cdot\boldsymbol{\upsilon}_2}{\sqrt{\lambda_2}} = \frac{1}{\sqrt{2}}\cdot\frac{R_1+R_2}{\sqrt{1+R_{12}}}.$$

This gives us:
$$R^2 = \frac{1}{1-R_{12}}\frac{(R_1-R_2)^2}{2} + \frac{1}{1+R_{12}}\frac{(R_1+R_2)^2}{2}.$$

This method gives an alternative expression for the coefficient of determination, where the terms now give distinct additive contributions. It is easy to show that this expression is equivalent to the one previously derived. ∎

Since $\boldsymbol{\Theta}$ is a principle submatrix of $\boldsymbol{\Phi}$ the eigenvalues and eigenvectors of the former are directly determined by the latter, and are subject to various notable constraints applying to this relationship, such as Cauchy's interlacing eigenvalue theorem (see e.g., Hwang 2004). Without going into detail on the particulars constraints involved in submatrix relations, the reader should note that the correlations $R_1, \ldots, R_m$ are constrained by $\lambda_1, \ldots, \lambda_m$ and $\boldsymbol{\upsilon}_1, \ldots, \boldsymbol{\upsilon}_m$ and so they should not be considered as free variables in the above formulae. The domain of these values would be determined by the relationships applying to the submatrix of a correlation matrix, and would require analysis of the spectral decomposition of matrices and their principal submatrices.

**EXAMPLE 3:** Continuing Examples 1-2 our design correlation matrix in this case has eigenvalue and eigenvector matrices given respectively by:

$$\Lambda = \begin{bmatrix} 1.5732 & 0 & 0 & 0 \\ 0 & 1.0721 & 0 & 0 \\ 0 & 0 & 0.9154 & 0 \\ 0 & 0 & 0 & 0.4393 \end{bmatrix},$$

and:

$$\upsilon = \begin{bmatrix} 0.6484 & -0.3077 & -0.2250 & 0.6590 \\ 0.4426 & 0.4947 & -0.6211 & 0.4426 \\ 0.5672 & -0.3551 & 0.4938 & 0.5672 \\ 0.2489 & 0.7311 & 0.5654 & 0.2489 \end{bmatrix}.$$

We have:

$$S_1 = \frac{\Omega \cdot \upsilon_1}{\sqrt{\lambda_1}} = -0.034009 \qquad S_2 = \frac{\Omega \cdot \upsilon_2}{\sqrt{\lambda_2}} = -0.118587,$$

$$S_3 = \frac{\Omega \cdot \upsilon_3}{\sqrt{\lambda_3}} = -0.351859 \qquad S_4 = \frac{\Omega \cdot \upsilon_4}{\sqrt{\lambda_4}} = 0.068407.$$

We therefore have:

$$R^2 = (-0.034009)^2 + (-0.118587)^2 + (-0.351859)^2 + 0.068407^2$$
$$= 0.001157 + 0.014063 + 0.123805 + 0.004679 = 0.1437.$$

This expansion gives the same result as in Example 2, and shows the contribution of each of the eigenvector components. We see that the third component contains almost all of the explanatory power of the set of explanatory vectors. ■

### 5. Augmentation of explanatory variables and "enhancement"

One important case of interaction between the input vectors in linear regression is presented in Hamilton (1987) where the author shows —contrary to intuition, and even to some statistics textbooks!— that it is possible to have $R^2 > \sum R_i^2$. Here the total information from the explanatory vectors is greater than the sum of the marginal parts. This can occur in cases where two explanatory variables act as "enhancers" for one another, adding a greater amount of explanatory power together than their marginal contributions when they are separate.[7]

---

[7] The reader should note that the literature on this topic is made slightly confusing by the fact that the same phenomenon is referred to either as "enhancement" or "suppression" by different authors, two terms which are *prima facie* incongruous. This terminology comes from the fact that enhancement of the explanatory power of another variable comes through the suppression of variance in intermediary quantities which lead to the enhancement (see e.g., Paulhus et al 2004). Hence, depending on whether

Cases of "enhancement" leading to this outcome are quite subtle. They can arise when an explanatory variable is only weakly correlated (or uncorrelated) with the response, but its correlation with other explanatory variables reduces the conditional variance of those variables substantially, making them more useful for prediction. As stated, this leads to a phenomena where the coefficient of variation for the multiple regression is greater than the sum of its parts, takes from simple linear regressions.

It is simple to compare the coefficient of determination in a multiple regression to the sum of the individual coefficients in simple linear regressions on the same variables, and this allows us to see that "enhancement" is possible in a wide range of cases. To do this we follow the approach of Routledge (1990) and Cuadras (1993) and express the difference in these quantities as:

$$\begin{aligned}\text{Difference} = R^2 - \sum R_i^2 &= \mathbf{\Omega}^T \mathbf{\Theta}^{-1} \mathbf{\Omega} - \|\mathbf{\Omega}\|^2 \\ &= \mathbf{\Omega}^T \mathbf{\Theta}^{-1} \mathbf{\Omega} - \mathbf{\Omega}^T \mathbf{\Omega} \\ &= \mathbf{\Omega}^T (\mathbf{\Theta}^{-1} - I) \mathbf{\Omega} \\ &= \mathbf{\Omega}^T (\mathbf{\upsilon} \mathbf{\Lambda}^{-1} \mathbf{\upsilon}^T - \mathbf{\upsilon}\mathbf{\upsilon}^T) \mathbf{\Omega} \\ &= (\mathbf{\upsilon}^T \mathbf{\Omega})^T (\mathbf{\Lambda}^{-1} - I)(\mathbf{\upsilon}^T \mathbf{\Omega}) \\ &= \sum_{k=1}^m \left(\frac{1}{\lambda_k} - 1\right)(\mathbf{\Omega} \cdot \mathbf{\upsilon}_k)^2 = \sum_{k=1}^m (1 - \lambda_k) S_k^2.\end{aligned}$$

We have already established the case where the explanatory variables are uncorrelated (i.e., no colinearity) which is the case where $\lambda_1 = \cdots = \lambda_m = 1$. In the other cases we have eigenvalues which are not all equal, and since the eigenvalues $\lambda_1 \geq \cdots \geq \lambda_m$ have unit average this means that the highest eigenvalue must be above unity and the lowest must be below unity. If the principal components corresponding to the low eigenvalues are strongly correlated with the response vector then the difference may be positive, corresponding to the case where the coefficient of determination is higher than the sum of the coefficients of determination for the individual regressions (i.e., enhancement occurs). Hadi and Ling (1998) note that this can be a good reason not to discard the lowest variance principal components; they offer this as a cautionary note on principal component analysis.

---

one is referring to the intermediary or final effect, the same phenomenon can be referred to by either of these two names. We choose the term "enhancement" in the present paper since it seems more sensible to refer to the final effect of interest.

**EXAMPLE 4:** Continuing Examples 1-3 we have:

$$\text{Difference} = (1 - 1.5732)(-0.034009)^2 + (1 - 1.0721)(-0.118587)^2$$
$$+ (1 - 0.9154)(-0.351859)^2 + (1 - 0.4393)0.068407^2$$
$$= -0.000663 - 0.001014 + 0.010479 + 0.002624 = 0.0114.$$

Since this value is positive we can see that "enhancement" occurs in this case. That is, we see that the coefficient of determination is higher than the sum of the coefficients for individual regressions on each of the explanatory variables! In this particular case the result is mostly due to the moderate negative correlation between the third principal component and the response vector. Consequently, the information in the set of explanatory variables is more than the sum of its marginal parts. ∎

There has been a sizable literature on this counter-intuitive phenomenon (see e.g., Hamilton 1987, Freund 1988, Hamilton 1988, Bertrand and Holder 1988, Mitra 1988, Cuadras 1993, Sharpe and Roberts 1997, Shieh 2001). One simple example which establishes this situation is when there are two explanatory vectors that are highly correlated, and a response vector within the span of the two vectors, but orthogonal to one, and almost orthogonal to the other (see e.g., Kendall and Stuart 1973, p. 359, Ex 27.22). In this case the coefficient of determination in the multiple regression is unity, but the coefficients of determination in the individual simple linear regressions on the same variables are (respectively) zero, and arbitrarily close to zero.

## 6. Conclusion

The present paper has looked at how to represent a multiple linear regression model in terms of the lengths of the input vectors and the angles between them (or equivalently, their correlation). This shows the relationship between the pairwise correlations of the input vectors to the multiple correlation and coefficient of determination. It also assists in obtaining a geometric understanding of multiple linear regression. Basic results in regression analysis can be presented in vector form and then rewritten in terms of the length and angles between regression vectors. Simple formulae for the ANOVA quantities emerge from this approach. The coefficient of determination can be considered either as a quadratic form using the inverse design correlation matrix, or as a sum of contributions from the principal components in the data.

The geometric representation of the coefficient of determination is extremely useful in training students to recognise the counter-intuitive properties of combining variables into multiple regression models. In particular, it can easily be seen that the coefficient of determination in multiple regression can be higher than the sum of coefficients in the individual regressions of the explanatory vectors (an enhancement effect), leading to a situation where the information obtained from the regression variables is greater than the sum of their marginal parts. (This is something that is so little known that some statistic textbooks incorrectly assert that it is impossible.) The difference between these quantities can be represented in terms of contributions from principal components and easily explained in geometric terms. Instructors may find this presentation useful for students who are trained in vector and matrix algebra. It allow students to obtain a geometric view of results that are usually presented in purely statistical terms.